\documentclass[copyright,creativecommons]{eptcs}

\providecommand{\event}{ACL2-2022}

\usepackage{alltt}
\usepackage{enumitem}
\usepackage[T1]{fontenc}
\usepackage{inconsolata}
\usepackage{underscore}
\usepackage{url}
\usepackage{xcolor}

\newcommand{\secref}[1]{Section \ref{#1}}

\newcommand{\citeman}[2]
{\cite[\href{http://acl2.org/manual?topic=#1}{\texttt{#2}}]
    {acl2-manual}}

\newcommand{\code}[1]{\texttt{#1}} 
\definecolor{commentgray}{gray}{0.4}
\newcommand{\ccode}[1]{\texttt{\textcolor{commentgray}{#1}}} 
\newenvironment{bcode} 
 {\begin{quote}\small\begin{alltt}}
 {\end{alltt}\end{quote}}

\newcommand{\lang}{\mathcal{L}}
\newcommand{\langLog}{\mathcal{A}}
\newcommand{\langProg}{\mathcal{P}}

\begin{document}


\title{A Proof-Generating C Code Generator for ACL2 \\
       Based on a Shallow Embedding of C in ACL2}

\author{Alessandro Coglio
        \institute{Kestrel Institute \\ \url{http://www.kestrel.edu}}}

\def\titlerunning{C Code Generation}
\def\authorrunning{A. Coglio}

\maketitle

\begin{abstract}
This paper describes a C code generator for ACL2
that recognizes ACL2 representations of C constructs,
according to a shallow embedding of C in ACL2,
and translates those representations to the represented C constructs.
The code generator also generates ACL2 theorems
asserting the correctness of the C code with respect to the ACL2 code.
The code generator currently supports a limited but growing subset of C
that already suffices for some interesting programs.
This paper also offers a general perspective on
language embedding and code generation.
\end{abstract}


\section{Introduction}
\label{intro}

Several theorem provers
(e.g.\ Coq, Isabelle, PVS)
include facilities to generate code in various programming languages
(e.g.\ C, Clean, Haskell, Lisp, Ocaml, Scala, Scheme, SML)
from executable subsets of the provers' logical languages.
That way, code written in a prover's language,
possibly verified to satisfy properties of interest,
can be run as code in a conventional programming language;
verifying the correctness of
the generated code with respect to the prover's code
is a separable problem, akin to compiler verification.
When carrying out
formal program synthesis by stepwise refinement
\cite{bmethod,zed,vdm,specware-www}
in a prover
to derive an implementation from a high-level specification,
a code generator can translate
the low-level (i.e.\ fully refined) specification
to the final program.

ACL2's tight integration with the underlying Lisp platform
may obviate the need for code generation,
because executable ACL2 code can run efficiently as Lisp.
An APT program derivation \cite{apt-www,apt-simplify,apt-isodata}
may end with an implementation in executable ACL2 that runs as Lisp.
However, some applications require code in other languages,
such as C for embedded systems or device drivers.
To synthesize this kind of code via an APT derivation,
a code generator for ACL2 can be used,
such as ATJ \cite{atj-deep,atj-shallow} \citeman{JAVA____ATJ}{java::atj}
(for Java).

A typical code generator for a prover can be viewed as
a reification of a shallow embedding of
the prover's (executable) language in the programming language:
constructs of the prover's language
are rendered as suitably equivalent constructs of the programming language.
The translation is centered on, and driven by, the prover's language,
which is mimicked in the programming language.
Therefore, unless the prover's language and the programming languages
are sufficiently similar,
the generated code may not be very efficient (in time and space) and idiomatic;
this could be an issue for the aforementioned example applications.

This paper proposes an approach
to turn the focus on the generated code,
and to exert direct control over it,
by flipping the direction of the embedding:
(i) a shallow embedding of the programming language in the prover's language
defines representations of program constructs in the prover; and
(ii) a code generator recognizes these representations
and translates them to the represented code.
This is \emph{code generation by inverse shallow embedding}:
the embedding is a (not necessarily reified) translation
from the programming language to the prover's language,
and the code generator is the (reified) inverse translation.
In contrast, \emph{code generation by direct shallow embedding}
is the more typical approach where
the code generator translates the prover's language to the programming language,
the translation being a shallow embedding of the former in the latter.

The program constructs shallowly embedded in the prover
are oriented towards the programming language,
and thus may not be idiomatic formulations in the prover's language.
This new code generation approach is designed
for program synthesis by stepwise refinement,
where the final refinement steps turn idiomatic code in the prover's language
into provably equivalent shallowly embedded program code.
These final refinement steps, carried out under user guidance,
afford fine-grained control on the exact form of the final program.

This new code generation approach is realized
in ATC (\textbf{A}CL2 \textbf{T}o \textbf{C}) \citeman{C____ATC}{c::atc},
the C code generator for ACL2 described in this paper.
ATC is designed for use with the aforementioned APT:
the final steps of an APT derivation turn ACL2 code
into C code shallowly embedded in ACL2,
which ATC turns into actual C code.
These final derivation steps take place within the ACL2 language,
and their verification involves only the ACL2 language;
these steps may be carried out via proof-generating APT transformations,
including ones tailored to ATC
that have been and are being developed at Kestrel.
These APT transformations are not discussed further in this paper,
which concentrates on ATC.

Besides the C code, ATC also generates ACL2 proofs
of the correctness of the C code with respect to the ACL2 code.
The proofs are based on a formalization of the needed subset of C in ACL2.
Together with the ACL2-to-ACL2 proofs in an APT derivation,
the ACL2-to-C proof generated by ATC
provides an end-to-end correctness proof of
the synthesized C code that ends the derivation
with respect to the high-level specification that starts the derivation.

ATC supports a limited subset of C18 \cite{c18},
which includes
integer types,
operations and conversions on them,
integer arrays with read and (destructive) write access,
local variable declarations and assignments,
loops of certain forms,
conditional expressions and statements,
and functions that may affect arrays.
Despite its limitations, this set suffices for some interesting programs.
ATC has been and is being used in a growing number of applications at Kestrel,
supporting the viability of the approach.

\secref{embeddings} offers a perhaps original perspective on
how different kinds of language embedding correspond to
different code generation approaches.
\secref{shallow-codegen} describes
the shallow embedding of C in ACL2 and how ATC uses it to generate C code.
\secref{deep-proofgen} describes the formalization (i.e.\ deep embedding)
of C in ACL2
and how ATC uses it to generate proofs.
\secref{future} discusses future work.
\secref{related} surveys related work.
In this paper, `fixtype' refers to a type
defined using the FTY library \citeman{ACL2____FTY}{fty}.


\section{Perspective on Language Embedding and Code Generation}
\label{embeddings}

An \emph{embedding} of a language $\lang$ in a language $\lang'$
is a representation of $\lang$ in $\lang'$.
In a \emph{deep} embedding,
the syntax of $\lang$ is represented
via $\lang'$ values that capture $\lang$ constructs,
and the semantics of $\lang$ is represented
via $\lang'$ operations over those values;
the representation is explicit.
In a \emph{shallow} embedding,
the syntax of $\lang$ is represented
via a translation of $\lang$ constructs to $\lang'$ constructs,
and the semantics of $\lang$ is represented
via the $\lang'$ semantics of
the $\lang'$ counterparts of the $\lang$ constructs;
the representation is implicit.%
\footnote{Although representations are conceivable
that blur the line between deep and shallow embedding,
the distinction remains generally useful,
and is commonly found in the literature
and used in technical discourse.}
An embedding may apply to a subset of $\lang$.
The features of an embedding depend on the nature of $\lang$ and $\lang'$,
in particular whether each one is a logical language or a programming language.

Given a logical language $\langLog$ (e.g.\ ACL2)
and a programming language $\langProg$ (e.g.\ C or Java),
there are four possible kinds of embedding, based on direction and depth:
(i) shallow embedding of $\langLog$ in $\langProg$;
(ii) deep embedding of $\langLog$ in $\langProg$;
(iii) shallow embedding of $\langProg$ in $\langLog$; and
(iv) deep embedding of $\langProg$ in $\langLog$.
Each kind corresponds to a different approach
to generate $\langProg$ code from $\langLog$,
particularly for the first three kinds,
with the fourth kind being a little different.
In addition, the fourth kind plays a role in
establishing the correctness of the generated code for all four approaches.


\subsection{Direct Shallow Embedding}
\label{direct-shallow}

A shallow embedding of $\langLog$ in $\langProg$
is a translation of constructs in (an executable subset of) $\langLog$
to suitably equivalent constructs in $\langProg$.
The translation can be used to generate $\langProg$ from $\langLog$.
The code generator reifies the embedding;
embedding and code generation go in the same direction---%
hence `direct'.

This is a typical code generation approach.
The constructs of $\langLog$ are rendered in $\langProg$:
the data types of $\langLog$ are mapped to data types of $\langProg$,
the operations of $\langLog$ are mapped to operations of $\langProg$,
and so on.
If $\langProg$ lacks suitable built-in counterparts
for certain $\langLog$ constructs,
such counterparts are defined in $\langProg$ (e.g.\ as libraries)
and used as code generation targets.

This translation is centered on, and driven by, $\langLog$,
which is mimicked in $\langProg$.
Unless $\langProg$ is sufficiently similar to $\langLog$,
the generated code may look more like ``$\langLog$ written in $\langProg$''
than like idiomatic $\langProg$;
it may not be as efficient (in time and space)
as the original $\langLog$ code
or as a handcrafted port to $\langProg$.
In some applications,
the readability of the generated code may be unimportant,
and its efficiency adequate;
but other applications, e.g.\ for resource-constrained systems,
may have more stringent requirements.

A realization of this code generation approach
is ATJ in shallow embedding mode \citeman{JAVA____ATJ}{java::atj},
which generates Java code from ACL2.
It relies on AIJ \citeman{JAVA____AIJ}{java::aij},
which implements ACL2 data types and operations in Java.
ACL2 functions are translated to Java static methods,
organized in Java classes corresponding to the ACL2 packages.
ACL2 \code{let} bindings are translated to
Java local variable declarations and assignments.
And so on; see the references for more details.


\subsection{Direct Deep Embedding}
\label{direct-deep}

A deep embedding of $\langLog$ in $\langProg$ is
an interpreter of (an executable subset of) $\langLog$ written in $\langProg$.
This enables a simple code generation approach:
the code generator translates $\langLog$ constructs
to their deeply embedded counterparts in $\langProg$,
which the interpreter executes in $\langProg$,
possibly via a thin wrapper produced by the code generator.
The interpreter reifies the embedding;
embedding and code generator go in the same direction---%
hence `direct'.

This is an unconventional code generation approach,
but is a conceptually simple way to run $\langLog$ in $\langProg$.
The interpreter includes representations of
the abstract syntax of $\langLog$,
the values and computation states of $\langLog$,
the basic operations of $\langLog$,
and so on.
The translation carried out by the code generator is straightforward.
The $\langProg$ code resulting from this code generation approach
is even less idiomatic and efficient than discussed in \secref{direct-shallow},
but it may be adequate for certain applications;
due to its simplicity, it is also fairly high-assurance
(relevant if formal proofs are absent).

A realization of this code generation approach
is ATJ in deep embedding mode \citeman{JAVA____ATJ}{java::atj},
which generates Java code from ACL2.
The interpreter is AIJ \citeman{JAVA____AIJ}{java::aij},
part of which is the Java implementation of the ACL2 data types and operations
mentioned in \secref{direct-shallow}.

This code generation approach could become more interesting
by accompanying it with partial evaluation.
According to the first Futamura projection \cite{parteval},
partially evaluating an interpreter with respect to a program
amounts to compiling the program
to the language that the interpreter is written in.
Thus, a partial evaluator for $\langProg$ can be used to
partially evaluate the $\langLog$ interpreter (written in $\langProg$)
with respect to (the $\langProg$ representation of) the $\langLog$ code.
For the aforementioned ATJ,
a partial evaluator for Java would be needed;
the partial evaluator may be written in any language, including ACL2.


\subsection{Inverse Shallow Embedding}
\label{inverse-shallow}

A shallow embedding of $\langProg$ in $\langLog$
is a translation of constructs in (a subset of) $\langProg$
to suitably equivalent constructs in $\langLog$.
The inverse of this translation can be used
to generate $\langProg$ from $\langLog$,
recognizing the $\langProg$ constructs shallowly embedded in $\langLog$
(i.e.\ recognizing the image of the embedding)
and turning them into the actual $\langProg$ constructs.
The code generator reifies the inverse of the embedding,
while the embedding does not have to be reified;
embedding and code generator go in opposite directions---%
hence `inverse'.

This appears to be a novel code generation approach.
It is centered on, and driven by, $\langProg$ rather than $\langLog$:
it enables users to specify the generated code exactly,
making it as idiomatic and efficient as desired,
and fit for the more demanding applications
mentioned in \secref{direct-shallow}.

While the generated $\langProg$ code can be idiomatic,
its $\langLog$ representation may look
more like ``$\langProg$ written in $\langLog$''
than like idiomatic $\langLog$ code,
and may be burdensome to write directly.
Thus, this code generation approach is designed
for program synthesis by stepwise refinement,
where the final refinement steps turn idiomatic $\langLog$
into the form required by the code generator,
i.e.\ into something in the image of the shallow embedding.
These refinement steps, carried out under user guidance,
afford fine-grained control on the exact form of the final program.
Since the code generator merely turns
$\langProg$ code shallowly embedded in $\langLog$ into actual $\langProg$ code,
the readability and the efficiency (in space and time) of the code
do not depend on the code generator:
they are the responsibility of the aforementioned final refinement steps;
shifting this responsibility from the code generator
to the stepwise refinement derivation
is a deliberate and distinctive aspect of this code generation approach.

A realization of this code generation approach is ATC, described in this paper,
which generates C code from ACL2.
ATC is designed for use with APT,
whose transformations (some tailored to ATC) can be used
to turn idiomatic ACL2 code into the form required by ATC.


\subsection{Inverse Deep Embedding}
\label{inverse-deep}

A deep embedding of $\langProg$ in $\langLog$ is
a formalization of (a subset of) $\langProg$ in $\langLog$.
Thus, it is the basis for formal correctness proofs of
$\langProg$ code generated from $\langLog$
via any of the three approaches described above,
playing a role alongside the three other kinds of embedding.
The formalization reifies the deep embedding.

Furthermore, this kind of embedding can be used
in a pop-refinement derivation \cite{popref}, where:
(i) the initial specification is
a predicate over $\langProg$ programs
that characterizes the possible implementations; and
(ii) the final specification is
a singleton predicate that selects one $\langProg$ implementation
in explicit syntactic form, from which the actual code is readily obtained.
All the specifications in the derivation, from initial to final,
are written in $\langLog$,
but their formulation requires a formalization of $\langProg$,
i.e.\ the deep embedding.
There is no code generator as such,
other than a simple obtainment of the implementation
from the final specification predicate where it is deeply embedded.
Embedding and refinement to code go in opposite directions---%
hence `inverse'.

For this code generation approach to be practical,
techniques must be developed to
make deeply embedded $\langProg$ constructs ``emerge'' within $\langLog$
via suitable refinement steps.
An idea worth pursuing is whether
the kind of code generator described in \secref{inverse-shallow}
could be realized as one or more such refinement steps,
to automatically turn shallowly embedded $\langProg$ constructs
into deeply embedded ones.

The proofs generated by ATC are based on a formalization of C in ACL2,
which is a deep embedding of C in ACL2.
ATC uses the code generation approach described in \secref{inverse-shallow},
not the pop-refinement approach sketched above.


\section{C Shallow Embedding and Code Generation}
\label{shallow-codegen}

ATC generates C code from ACL2
according to the inverse shallow embedding approach
explained in \secref{inverse-shallow}.
ATC relies on the definition of a shallow embedding of C in ACL2,
whose image ATC recognizes in ACL2 code and translates to actual C code.


\subsection{Shallow Embedding of C in ACL2}
\label{shallow}

The shallow embedding of C in ACL2 relied on by ATC consists of
(i) an ACL2 model of C integers and arrays and
(ii) a definition of how C expressions, statements, and functions
are represented as ACL2 terms and functions.
The latter is embodied in the ATC user documentation,
which spells out the representation in an almost formal way.
There is currently no implemented translation from C to ACL2.


\subsubsection{Integers}

The model includes all the C18 standard integer types except \code{\_Bool},
namely
\code{char}, \code{short}, \code{int}, \code{long}, and \code{long} \code{long},
both \code{signed} (the default) and \code{unsigned}
(but not the plain \code{char} type);
10 types in total.
The model includes
unary integer operations
(\code{+}, \code{-}, \verb|~|, \code{!}; 4 in total),
binary integer operations
(\code{+}, \code{-}, \code{*}, \code{/}, \code{\%},
\code{\&}, \code{|}, \code{\^}, \code{<{}<}, \code{>{}>},
\code{<}, \code{>}, \code{<=}, \code{>=}, \code{==}, \code{!=};
16 in total),%
\footnote{The non-strict operations \code{\&\&} and \code{||}
are represented differently, as described later.}
and integer type conversions (90 in total).

The exact format of the C integer types is implementation-dependent.
The model assumes two's complement without padding bits (as prevalent),
and has ACL2 nullary functions for the sizes of the different types;
these nullary functions are referenced in other parts of the model
to avoid hardwiring assumptions on sizes,
which vary across popular C implementations.
These nullary functions are currently defined (to common values),
but they are planned to be made constrained instead
(see \secref{future}).
The C integer values are represented as ACL2 integers in appropriate ranges
(defined via the nullary functions),
tagged with an indication of their type.
The model has a fixtype for each C integer type.

The C integer operations and conversions are modeled as ACL2 functions,
whose guards capture the conditions under which
the results are well-defined in C18.
For example, \code{int} addition is modeled as follows:
\begin{bcode}
(defun add-sint-sint (x y) \ccode{; addition of signed int and signed int}
  (declare (xargs :guard (and (sintp x) (sintp y) (add-sint-sint-okp x y))))
  (sint (+ (sint->get x) (sint->get y))))
(defun add-sint-sint-okp (x y) \ccode{; well-definedness condition}
  (declare (xargs :guard (and (sintp x) (sintp y))))
  (sint-integerp (+ (sint->get x) (sint->get y))))
\end{bcode}
Two \code{int} values, recognized by \code{sintp},
are added by taking the underlying ACL2 integers via \code{sint->get},
adding them via ACL2's addition \code{+},
and turning the result into an \code{int} value via \code{sint},
provided that the result is representable as an \code{int},
whose range is recognized by \code{sint-integerp}.%
\footnote{In C18,
signed integer arithmetic operations are well-defined only if
the true result of the operation can be represented
in the type of the result of the operation
(which is determined from the types of the operands).
C implementations may extend this well-definedness,
e.g.\ as two's complement wrap-around.}

C unary and binary integer operations apply to any combination of operand types
and have fairly complex rules about operand type conversions and result types;
the rules are not uniform across the different types.
The model has versions of the integer operations
for all possible combinations of operand types,
whose definitions capture those rules.
There are thousands of such functions,%
\footnote{(4 unary operations $\times$ 10 types)
$+$ (16 binary operations $\times$ 100 type combinations)
$=$ 1,640 functions.}
generated via macros that exploit their available uniformity.


\subsubsection{Arrays}

The model includes monodimensional arrays of the above integer types,
with read and write access.

A C array is modeled as a non-empty list of C integer values,
tagged with an indication of its type.
The model has a fixtype for each C array integer type.

Read and write access is modeled via ACL2 functions to read and write arrays.
The read functions return an array element from an array and an index;
the write functions return a new array
from an old array, an index, and a value for the element.
These functions have guards requiring the index to be within the array bounds.
Since an array may be indexed with any integer type in C,
there are versions of these functions
for all combinations of array and index types.
There are hundreds of such functions,%
\footnote{(2 read or write operations)
$\times$ (10 array types)
$\times$ (10 index types)
$=$ 200 functions.}
generated via macros that exploit their uniformity.

Arrays are not first-class entities in C,
but mere juxtapositions of their elements.
Although the model just described
appears to treat arrays as first-class entities,
their use in the shallowly embedded C code
is restricted in ways that make this modeling adequate;
see also \secref{deep}.


\subsubsection{Expressions, Statements, and Functions}

C functions are represented as ACL2 functions
that operate on (the ACL2 model of) C values.
Consider the following C function:
\begin{bcode}
int f(int x, int y, int z) \{
    return (x + y) * (z - 3);
\}
\end{bcode}
This is represented as follows in ACL2
(where the ellipsis in the guard is explained later):
\begin{bcode}
(defun |f| (|x| |y| |z|)
  (declare (xargs :guard (and (sintp |x|) (sintp |y|) (sintp |z|) ...)))
  (mul-sint-sint (add-sint-sint |x| |y|)
                 (sub-sint-sint |z| (sint-dec-const 3))))
\end{bcode}
The reason for the vertical bars around the ACL2 symbols is that
a C identifier is represented as
an ACL2 symbol whose \code{symbol-name} is the identifier;
if \code{f} were used as the ACL2 function name instead of \code{|f|},
it would represent a C function named \code{F}.
The correspondence between the body of \code{|f|}
and the return expression of \code{f} is clear:
the ACL2 functions that model arithmetic operations
have been explained earlier;
the \code{int} decimal (i.e.\ in base 10) constant \code{3}
is represented via \code{sint-dec-const}, which is part of the model.
The input and output types of \code{f} are represented by
the corresponding guard conjuncts of \code{|f|}
and the fact that the body of \code{|f|}
returns \code{sintp} (via \code{mul-sint-sint});
see also \secref{codegen}.

Besides C expressions as exemplified above,
ACL2 terms of certain forms represent C statements,
which may contain blocks with local variable declarations.
Consider the following C function:
\begin{bcode}
unsigned int g(unsigned int x, unsigned int y) \{
    unsigned int z = 1U;
    if (x < y) \{ z = z + x; \} else \{ z = z + y; \}
    return 2U * z;
\}
\end{bcode}
This is represented as follows in ACL2:
\begin{bcode}
(defun |g| (|x| |y|)
  (declare (xargs :guard (and (uintp |x|) (uintp |y|))))
  (let ((|z| (declar (uint-dec-const 1))))
    (let ((|z| (if (boolean-from-sint (lt-uint-uint |x| |y|))
                   (let ((|z| (assign (add-uint-uint |z| |x|))))
                     |z|)
                 (let ((|z| (assign (add-uint-uint |z| |y|))))
                   |z|))))
      (mul-uint-uint (uint-dec-const 2) |z|))))
\end{bcode}
Local variable declarations and assignments
are both represented via \code{let} bindings;
the two cases are distinguished by
the \code{declar} and \code{assign} wrappers,
which are defined as identity functions.
Statements, like the \code{if} statement above,
that affect variables and are followed by more code
are represented via \code{let} bindings without any wrappers;
both branches of the \code{if} term in ACL2
must end with (the latest values of) the affected variables.
If multiple variables were affected,
\code{mv-let} would be used to bind them to the \code{if} term,
whose branches would have to end with an \code{mv} of the variables.
Since ACL2 is functional, variable updates are explicated.
Function \code{boolean-from-sint} converts
the \code{int} returned by the less-than operation to a boolean,
as needed for \code{if} tests in ACL2;
the conversion is not reflected in the C code,
which uses scalars (including integers) in \code{if} tests.

C loops are represented as tail-recursive ACL2 functions.
Consider the following modular factorial:
\begin{bcode}
unsigned int h(unsigned int n) \{
    unsigned int r = 1U;
    while (n != 0U) \{ r = r * n; n = n - 1U; \}
    return r;
\}
\end{bcode}
This is represented as follows in ACL2:
\begin{bcode}
(defun |h\$loop| (|n| |r|) \ccode{; representation of the loop of h}
  (declare (xargs :guard (and (uintp |n|) (uintp |r|))))
  (if (boolean-from-sint (ne-uint-uint |n| (uint-dec-const 0)))
      (let* ((|r| (assign (mul-uint-uint |r| |n|)))
             (|n| (assign (sub-uint-uint |n| (uint-dec-const 1)))))
        (|h\$loop| |n| |r|))
    (mv |n| |r|)))
(defun |h| (|n|) \ccode{; representation of the function h}
  (declare (xargs :guard (uintp |n|)))
  (let ((|r| (declar (uint-dec-const 1))))
    (mv-let (|n| |r|)
      (|h\$loop| |n| |r|)
      (declare (ignore |n|)) \ccode{; because n is not used after the loop}
      |r|)))
\end{bcode}
Since the loop function does not represent a C function,
its name does not have to be a legal C identifier.
In \code{|h|},
the two variables are \code{mv-let}-bound to the loop function call,
because the loop affects both.
Loop functions may have a more elaborate structure than above,
but every control path must end
either with a recursive call on the formal parameters
(which must therefore be updated before the call)
or with (the subset of) the formal parameters affected by the loop.

C arrays are passed around as wholes in functional ACL2,
but they are passed around as pointers in C.
The ACL2 representation is correct so long as the arrays are treated in
a stobj-like \citeman{ACL2____STOBJ}{stobj} single-threaded way,
which is required in this shallowly embedded representation.
Consider the following C function:
\begin{bcode}
void i(unsigned char *a, int x, int y) \{
    a[x] = (unsigned char) 1;
    a[y] = (unsigned char) 2;
\}
\end{bcode}
This is represented as follows in ACL2
(where the ellipsis in the guard is explained later):
\begin{bcode}
(defun |i| (|a| |x| |y|)
  (declare (xargs :guard (and (uchar-arrayp |a|) (sintp |x|) (sintp |y|) ...)))
  (let* ((|a| (uchar-array-write-sint |a| |x| (uchar-from-sint (sint-dec-const 1))))
         (|a| (uchar-array-write-sint |a| |y| (uchar-from-sint (sint-dec-const 2)))))
    |a|))
\end{bcode}
Array writes are represented as \code{let} bindings
of the array variables to calls of the array write functions.
The C function returns nothing (i.e.\ \code{void}),
but since it affects the array,
the ACL2 function returns the updated array.
If multiple arrays were updated, they would all be returned via \code{mv},
while the C function would still return nothing.
If the C function returned a result besides affecting arrays,
the ACL2 function would return the result and the updated arrays via \code{mv}.
Arrays may be updated in loops;
in general, ACL2 loop functions return affected variables and arrays,
while ACL2 non-loop functions return affected arrays and optionally a result.

The non-strict C operations \code{\&\&} and \code{||}
are represented via ACL2's \code{and} and \code{or} macros,
which are non-strict because they are defined in terms of \code{if}.
Representing \code{\&\&} and \code{||} via strict ACL2 functions
would require the evaluation of their second operand to be well-defined
(i.e.\ require its guards in ACL2 to be verified)
regardless of the value of the first operand,
which would be too restrictive,
preventing many valid and useful C programs from being represented.
Functions like \code{boolean-from-sint} shown earlier,
as well as inverses like \code{sint-from-boolean}
that are part of the shallow embedding,
are used to convert between ACL2 booleans and C integers
in the representation of C expressions involving \code{\&\&} and \code{||}.

The above is just an overview of the supported representations of C in ACL2.
The ATC reference documentation \citeman{C____ATC}{c::atc}
spells out the supported representations in full detail.


\subsection{C Code Generation}
\label{codegen}

ATC is invoked on a list of ACL2 functions
that represent C functions and C loops.
If an ACL2 function is not recursive, it represents a C function;
if it is recursive, it represents a C loop.
The ACL2 functions must be listed in bottom-up order
(i.e.\ each function may call the preceding ones or itself,
but not the subsequent ones);
the C functions are generated in the same order,
currently in a single \code{.c} file.
The above requirements imply that
recursive C functions cannot be generated currently.

ATC requires all the ACL2 functions
to be defined, in logic mode, and guard-verified.
This is critical for proof generation (see \secref{proofgen}).
The fact that the loop functions are in logic mode means that
the loops must terminate under the guards,
which may be added to the termination conditions via \code{mbt},
which ATC ignores as far as the representation of C code goes.
The fact that the functions are guard-verified means that
all arrays are always accessed within their bounds
(avoiding well-known problems in C),
and that all arithmetic operations always have well-defined results
according to C18.
The ellipses in the guards of some of the examples in \secref{shallow}
include conditions needed for guard verification,
namely that the parameters of \code{|f|} are in certain ranges
and that the index parameters of \code{|i|}
are non-negative and less than the array's length.%
\footnote{No additional conditions are needed
in the guards of \code{|g|} and \code{|h|} (and \code{|h\$loop|}),
because unsigned arithmetic
is always well-defined as wrap-around in C18.}

ATC checks that the names of the non-loop functions,
and the names of the parameters of both loop and non-loop functions,
are valid C ASCII identifiers.
The input types of the C functions are determined from the guards,
which must include explicit conjuncts
like the ones in the examples in \secref{shallow}.
ATC operates on the unnormalized bodies of the ACL2 functions
\citeman{ACL2____FUNCTION-DEFINEDNESS}{function-definedness}:
it checks that they contain supported representations of C code,
i.e.\ that they are in the image of the shallow embedding.
In the process, ATC performs C-like type checking,
to determine the types of \code{let}-bound variables with \code{declar},
to determine the output types of the C functions,
and to ensure that the generated C code is acceptable to C compilers;
the latter is not always ensured by guard verification,
specifically when there is dead code under the guards (by user mistake),
which C compilers do not regard as dead.

While this translation of ACL2 to C is conceptually relatively simple,
its implementation is more complicated than anticipated.
There are a lot of detailed cases to consider and to check.
Giving informative error messages when the checks fail takes some effort;
the current implementation may be improved in this respect.
The ATC code responsible for the checks and translation
consists of a few thousand lines (including documentation and blanks).

ATC generates C code
via an abstract syntax of (a sufficient subset of) C,
defined via algebraic fixtypes.
The C file is generated via a pretty-printer,
which minimizes parentheses in expressions
by considering the relative precedence of the C expression constructs.

ATC generates the C code, as well as the associated ACL2 proof events,
very quickly for all the examples tried so far
(which are admittedly not very large).
This is expected, as ATC does a linear pass on the ACL2 code
that does not perform particularly intensive computations.
However, the processing of the proof events by ACL2 is not always quick;
see \secref{proofgen}.


\section{C Deep Embedding and Proof Generation}
\label{deep-proofgen}

Besides C code,
ATC also generates ACL2 theorems asserting
the correctness of the C code with respect to the ACL2 code.
These assertions rely on a formalization in ACL2 of
the syntax and semantics of (a sufficient subset of) C,
i.e.\ a deep embedding of C in ACL2,
which plays a role alongside the shallow embedding
(see \secref{inverse-deep}).
This formalization is more general than ATC, and is of independent interest.


\subsection{Deep Embedding of C in ACL2}
\label{deep}

The formalization of C in ACL2 consists of
(i) an abstract syntax,
(ii) a static semantics, and
(iii) a dynamic semantics.
Both static and dynamic semantics are defined over the abstract syntax.


\subsubsection{Syntax}

The abstract syntax is currently the same one
used for code generation (see \secref{codegen}).
It captures the syntax of C after preprocessing.

Future versions of ATC may likely generate
C code with at least some preprocessing directives,
which would be convenient to incorporate
into the abstract syntax used for code generation;
in this case, the abstract syntax would capture
the syntax of C before preprocessing.
Thus, at some point it may be appropriate for the C formalization
to use its own separate abstract syntax,
and in fact to have
one for the C syntax before preprocessing
and one for the C syntax after preprocessing,
to model more faithfully C18's translation phases.

Currently the formalization does not include the C concrete syntax.
Since ATC's generated proofs currently apply to abstract syntax,
there is no immediate need to formalize concrete syntax.


\subsubsection{Static Semantics}

The static semantics of C
consists of decidable requirements
that must be satisfied by C code to be compiled and executed:
every referenced variable and function is in scope;
every operation is applied to operands of appropriate types;
and so on.
These are described informally in the C18 standard.

These requirements are formalized via (executable) ACL2 code
that checks whether the C abstract syntax satisfies those requirements,
analogously to what C compilers do.
The checking code makes use of symbol tables,
i.e.\ data structures that capture which C symbols
(e.g.\ function and variable names) are in scope
and what their types are.
If an abstract syntax entity violates any requirement,
the checking code returns an error result;
otherwise, it returns a non-error result
that may include inferred information about the checked abstract syntax entity.
In particular,
the successful checking of an expression yields
the (non-\code{void}) C type of the expression,
and the successful checking of a statement yields
a non-empty finite set of C types,
corresponding to the possible values that may be returned
(including \code{void} for code that completes execution
without a \code{return} or with one without expression);
the latter set is used to check that the body of a function
always returns something consistent with the function's return type.


\subsubsection{Dynamic Semantics}

The dynamic semantics of C
is the execution behavior of C code
(that satisfies the static semantic requirements),
i.e.\ how the execution of expressions, statements, etc.\
manipulates values and memory.
While this behavior is normally realized
by compiling C code to machine code and running the latter,
it can be described, as the C18 standard does,
in terms of an abstract machine that directly executes C.

This abstract machine is formalized
via fixtypes that capture the machine states
and via (executable) ACL2 code that manipulates machine states
according to the C expressions, statements, etc.

The model of machine states starts with the model of C integers and arrays
described in \secref{shallow}, which is shared by deep and shallow embedding.
C values are defined to consist of integers and pointers (for arrays);
a pointer consists of a type and an optional address (absent for null pointers),
where an address is a natural number that is treated opaquely.%
\footnote{These addresses are just used to identify arrays.
They do not represent actual addresses in memory.
The formalization performs no arithmetic on them.
Other kinds of entities, e.g.\ strings, could be used instead.}
Importantly, in this model the C values carry information about their types,
which is used in the defensive checks described later.
The state of the variables in scope is
a stack (i.e.\ list) of finite maps from identifiers to values:
the stack corresponds to the nested C scopes,
and each finite map consists of the variables in the same scope.
A frame consists of a function name and
a stack of variable scopes of the kind just described.
A computation state consists of
a stack (i.e.\ list) of frames, which captures the call stack,
and a heap, which is a finite map
from addresses (the ones used in non-null pointers)
to arrays (the ones shared with the shallow embedding).
The current model of the heap is simple,
with arrays treated as wholes and accessed exclusively via their addresses.

The next component of the formalized dynamic semantics of C consists of
ACL2 functions that perform basic operations on computation states.
These include operations to:
push and pop frames, and get the top frame;
push and pop scopes (in the top frame);
create variables (in the top scope in the top frame),
and read and write variables
(in some scope, looked up from innermost to outermost, in the top frame);
read and write arrays (in the heap).
There are no operations to create or destroy arrays;
the read and write operations apply to externally created arrays.

The C functions in scope are captured via a function environment,
which is a finite map from identifiers (function names)
to information about the function (typed parameters, return type, and body).
The function environment for a C program is built
by collecting all the C functions that form the program;
it never changes during execution.

The C18 standard does not prescribe the order of expression evaluation;
since C expressions may have side effects,
different orders of evaluation may lead to different outcomes,
which complicates formal modeling.
The formal model manages this complexity
by partitioning expressions into pure ones (i.e.\ free of side effects)
and non-pure ones (i.e.\ with possible side effects).
While the pure ones may be freely nested
(because their order of evaluation does not matter),
the non-pure ones may only appear in certain positions of the code
that force a unique order of evaluation:
specifically,
assignments may only appear as expression statements.
These restrictions, which are not required in C18,
are enforced in the subset of C covered by the formal model.

The last component of the formalized dynamic semantics of C consists of
ACL2 functions to execute expressions, statements, etc.
These functions are defensive,
i.e.\ they do not assume that
the code satisfies the static semantic requirements,
and instead independently check those requirements on the dynamic data,
i.e.\ that a referenced variable is in the current frame,
that the values that an operation is applied to have appropriate types
(recall the note above about values carrying type information),
and so on;
if any of these checks fails,
the execution functions return an error indication.%
\footnote{It remains to be formally proved that
the static semantics is sound with respect to the dynamic semantics,
namely that no such error indications are returned
when executing code that satisfies the static requirements.
The proofs generated by ATC do not rely on this property,
but its proof would provide
a major validation of this formalization of C in ACL2.}
The ACL2 execution functions also check that
the results of integer operations and conversions are well-defined
and that arrays are read and written within their bounds;
if any of these checks fails,
the execution functions return an error indication.
Importantly, this means that the dynamic semantics returns an error indication
if the C code attempts an unsafe array access.

The ACL2 function to execute a pure expression
takes an expression and a computation state as inputs,
and returns either a value or an error indication as output.
Executing a non-pure expression may involve executing a function call,
which involves executing the statements in the function,
and so on.
Thus, the ACL2 functions to execute
non-pure expressions, statements, functions, etc.\
are mutually recursive.
Besides the syntactic entities they execute (statements etc.),
these functions take a computation state and a function environment as inputs
(the latter is used to look up called functions),
and they return a possibly updated computation state as output,
along with a result (e.g.\ an optional value for a statement)
or an error indication of the kind explained above.

These mutually recursive functions,
together with the functions to execute pure expressions and lists thereof,
formalize an (interpretive) big-step operational semantics:
each function executes its syntactic construct to completion,
by recursively executing the sub-constructs
and combining their outcomes to yield the final outcome.%
\footnote{In contrast, a small-step operational semantics
would only execute part of a construct at each step.
This requires keeping track of
which parts of a construct have been already executed
and which parts must be executed next.
This is more complicated than just executing each construct to completion.
An example of small-step operational semantics,
for the ACL2 programming language (but still illustrating the point),
is in the Community Books
\citeman{ACL2PL____ACL2-PROGRAMMING-LANGUAGE}
        {acl2pl::acl2-programming-language}.}
Since C code may not terminate in general,
the mutual recursion may not terminate,
which makes it problematic to define in a theorem prover like ACL2.
This is solved by adding an artificial counter
that limits the depth of the mutual recursion:
the counter is decremented by 1 at each recursive call,
and used as the measure of the mutual recursion,
whose termination proof is thus straightforward.
This does not sweep issues of (non-)termination under the rug:
see \secref{proofgen}.

The ACL2 execution functions
for (pure and non-pure) expressions, statements, etc.\
use the operations and conversions on integers discussed in \secref{shallow},
which are shared between deep and shallow embedding.
They also use the operations on computation states discussed above.


\subsection{C Proof Generation}
\label{proofgen}

Although ATC's translation
from the shallowly embedded C code in ACL2 to the actual C code
is conceptually simple by design,
generating ACL2 proofs of their equivalence
is more laborious than anticipated.
These proofs consist of ACL2 events that ATC builds
and then submits to ACL2 via \code{make-event}.


\subsubsection{Code Constant}

As ATC computes the abstract syntax of the generated C code
and pretty-prints it to a \code{.c} file (see \secref{codegen}),
it also generates a \code{defconst} event that defines a named constant
for the C code, which is a single translation unit in C18 terminology:%
\footnote{In the future, this may be generalized to
a collection of translation units that forms a generated program.
Each translation unit is in a \code{.c} or \code{.h} file.}
\begin{bcode}
(defconst *program* \ccode{; default name, customizable by the user}
    '(:transunit ...) \ccode{; fixtype value of translation unit}
\end{bcode}
The other generated events refer to this constant
to assert properties of the generated C code.


\subsubsection{Static Correctness}

ATC generates a \code{defthm} event asserting that
the top-level checking function of the C static semantics
succeeds on the named constant described above:
\begin{bcode}
(defthm *program*-well-formed
    (equal (check-transunit *program*) \ccode{; top-level static checking function}
           :wellformed))
\end{bcode}
Since this is a ground theorem, it is proved by execution:
ATC generates hints to prove it in the theory consisting of
exactly the executable counterpart of that top-level checking function.
This proof is very quick for all the examples tried so far.

This establishes that the C code satisfies all the static semantic requirements,
and can be therefore successfully compiled by C compilers.
This property is not always implied by the dynamic correctness theorems,
in the presence of dead code under guards
as briefly discussed in \secref{codegen}.


\subsubsection{Dynamic Correctness}

ATC generates \code{defthm} events asserting that, roughly speaking,
executing each C function according to the dynamic semantics
yields the same outcome as the representing ACL2 function.
The proofs are carried out via a symbolic execution of the C code
(which is constant in the proofs, i.e.\ \code{*program*})
that turns the execution functions applied to the deeply embedded C code
into a form that can be matched with the shallowly embedded C code
that forms the ACL2 functions.

The general formulation is illustrated by
the theorem for function \code{|f|} in \secref{shallow}:
\begin{bcode}
(defthm *program*-|f|-correct
  (implies (and (compustatep compst)
                (equal fenv (init-fun-env *program*))
                (integerp limit)
                (>= limit ...) \ccode{; constant lower bound calculated by ATC}
                (and (sintp |x|) (sintp |y|) (sintp |z|) ...)) \ccode{; guard of |f|}
           (equal (exec-fun (ident "f") (list |x| |y| |z|) compst fenv limit)
                  (b* ((result (|f| |x| |y| |z|)))
                    (mv result compst)))))
\end{bcode}
It says that executing the C function whose name is the identifier \code{f}
on \code{int} inputs \code{|x|}, \code{|y|}, \code{|z|}
satisfying the guard of \code{|f|},
on an arbitrary computation state \code{compst},
with the function environment \code{fenv} for the program,
with a sufficiently large recursion limit \code{limit},
yields the same result (value) as \code{|f|} on the same inputs,
and does not change the computation state.
The hypothesis on the limit ensures that
execution does not stop prematurely;
the lower bound, which is constant in this case,
is calculated by ATC as it generates function \code{f},
because it knows how much recursion depth the execution of its body needs.
ATC generates hints (not shown) to prove this theorem,
mainly consisting of a large theory
for symbolically executing the C code in ACL2,
along with a lemma instance for the guard theorem of \code{|f|},
which steers the symbolic execution away from returning error indications
due to non-well-defined arithmetic operations,
since they are all well-defined under the guard.
The fact that \code{|f|} never returns error indications
and that the execution result of \code{f} is equal to that,
means that execution never returns error indications:
in general, this means that the C code generated by ATC
always has a well-defined behavior, including always accessing arrays safely.
This theorem also implicitly asserts that \code{f} terminates:
it is always possible to find a value of \code{limit}
that satisfies the inequality hypothesis.

To prove the theorem above, ATC also generates a local theorem%
\footnote{Here `local' refers to the \code{encapsulate}
with the ATC-generated events,
only a few of which are exported, free of hints.}
saying that \code{|f|} returns an \code{int};
this local theorem is referenced in the generated hints for the above theorem.
An analogous local theorem is generated by ATC for each ACL2 function,
with the applicable type(s).

ATC generates a correctness theorem for each C function.
If the C function affects arrays, the formulation is more complicated.
This is illustrated by the theorem for function \code{|i|} in \secref{shallow}:
\begin{bcode}
(defthm *program*-|i|-correct
  (b* ((|a| (read-array (pointer->address |a|-ptr) compst)))
    (implies (and (compustatep compst)
                  (equal fenv (init-fun-env *program*))
                  (integerp limit)
                  (>= limit ...) \ccode{; constant lower bound calculated by ATC}
                  (pointerp |a|-ptr)
                  (not (pointer-nullp |a|-ptr))
                  (equal (pointer->reftype |a|-ptr) (type-uchar))
                  (and (uchar-arrayp |a|) ...)) \ccode{; guard of |i|}
    (equal (exec-fun (ident "i") (list |a|-ptr |x| |y|) compst fenv limit)
           (b* ((|a|-new (|i| |a| |x| |y|)))
             (mv nil (write-array (pointer->address |a|-ptr) |a|-new compst))))))
\end{bcode}
Since C arrays are manipulated as wholes in ACL2 but via pointers in C,
a variable \code{|a|-ptr} is introduced for the pointer to the array,
while the array \code{|a|} is the result of \code{read-array}:
the call of \code{exec-fun} takes \code{|a|-ptr};
the call of \code{|i|} takes \code{|a|}.
The hypotheses on \code{|a|-ptr} say that it is
a non-null pointer of the right type;
the guard hypothesis saying that \code{|a|} is an array
also implicitly says that \code{|a|-ptr} points to an existing array,
because \code{read-array} returns an error indication if that is not the case.
This hypothesis constrains the computation state to contain that array,
which is thus no longer unconstrained
as in the correctness theorem for \code{|f|}.
Also unlike the correctness theorem for \code{|f|},
here the computation state is updated by the execution of \code{i}:
the new array, returned by \code{|i|}, is \code{|a|-new};
the new computation state is obtained by replacing
the array \code{|a|} pointed by \code{|a|-ptr} with \code{|a|-new}.
The \code{nil} that precedes the updated computation state
refers to the fact that \code{i} returns no result.
The generated hints for this theorem are similar to
the ones for the correctness theorem for \code{|f|}.

ATC also generates a correctness theorem for each C loop,
relating its execution to the ACL2 function that represents it.
This is illustrated by
the theorem for function \code{|h\$loop|} in \secref{shallow}:
\begin{bcode}
(defthm *program*-|h\$loop|-correct
  (b* ((|n| (read-var (ident "n") compst))
       (|r| (read-var (ident "r") compst)))
    (implies (and (compustatep compst)
                  (not (equal (compustate-frames-number compst) 0))
                  (equal fenv (init-fun-env *program*))
                  (integerp limit)
                  (>= limit ...) \ccode{; non-constant lower bound calculated by ATC}
                  (and (uintp |n|) (uintp |r|))) \ccode{; guard of |h\$loop|}
             (equal (exec-stmt-while
                     '... \ccode{; test of the loop}
                     '... \ccode{; body of the loop}
                     compst fenv limit)
                    (b* (((mv |n|-new |r|-new) (|h\$loop| |n| |r|)))
                      (mv nil
                          (write-var (ident "n")
                                     |n|-new
                                     (write-var (ident "r")
                                                |r|-new
                                                compst))))))))
\end{bcode}
The computation state is constrained to have at least one frame
(i.e.\ the number of frames is not 0)
and two variables \code{n} and \code{r} in the top frame
(i.e.\ the variables accessed by the loop);
\code{|n|} and \code{|r|} are bound to those variables' values.
The guard hypothesis saying that \code{|n|} and \code{|r|}
are \code{unsigned} \code{int} values
also implicitly says that the variables exist,
because \code{read-var} returns an error indication if that is not the case.
The lower bound in the hypothesis on \code{limit} is not constant here:
it is a symbolic term that references the measure of \code{|h\$loop|},
because the measure is related to the recursion limit needed to execute the loop;
ATC calculates this symbolic term as it generates the loop's code.
The \code{exec-stmt-while} function is called on
the quoted test and body of the loop,
making this theorem applicable as a rewrite rule
in the proof of the correctness theorem for \code{|h|}
(more on this below).
Since C loops affect variables (and possibly arrays) but do not return results,
the right side of the equality whose left side is the \code{exec-stmt} call
has a form similar to the correctness theorem for function \code{|i|} above,
except that variables instead of arrays are updated in the computation state.
This theorem is proved by induction;
the generated hints make use of the termination theorem of \code{|h\$loop|},
as well as of some local functions and theorems not discussed here.

If a C loop affects arrays besides variables,
its correctness theorem combines characteristics of
the correctness theorems for \code{|h\$loop|} and for \code{|i|}.
The theorem starts with \code{b*} bindings for both variables and arrays.
Variables for the pointers to the arrays are introduced.
The final computation state updates both variables and arrays,
via nested \code{write-var} and \code{write-array} calls.

If a C function or C loop affects multiple arrays,
the generated correctness theorem includes hypotheses saying that
the arrays are all at different addresses.
Since the representing ACL2 function treats the arrays as wholes,
updating one leaves the others unchanged in the ACL2 representation.
In the C code, where the arrays are handled via pointers,
updates would not be independent if two pointers pointed to the same array.
The absence of aliasing is therefore
a critical hypothesis in the correctness theorems.

The correctness theorem for a C loop is used as a rewrite rule
in the symbolic execution of
the correctness theorem for the C function that contains the loop.
Similarly, the correctness theorem for a C function is used as a rewrite rule
in the symbolic execution of another C function that calls it.
In other words,
the generated correctness proofs build upon each other,
according to the call graph of the ACL2 functions
that represent the C functions and loops.
For this to work,
the theorems are formulated so that
the left sides of the rewrite rules match
the terms that arise during symbolic execution.
In particular, the recursion limit is a variable, \code{limit},
which the hypotheses constrain with a lower bound:
this ensures that any actual limit term
that arises during the symbolic expression is matched by the variable,
and that the inequality hypothesis can be discharged.
ATC calculates the lower bound terms by considering
not only the code generated for the C function or loop,
but also the lower bound terms previously calculated
for all the C functions and loops that may be executed;
the more complex the call graph of the ACL2 functions,
the more complex the resulting lower bound terms.

The above is just an overview of how ATC generates dynamic correctness proofs.
There are several other complexities involved,
such as canonical forms of the symbolic computation states,
which are defined via slightly modified operations on computation states,
and achieved via appropriate sets of rewrite rules.
The implementation of ATC includes documentation
for all the details of proof generation.

Some correctness theorems are processed quickly by ACL2,
e.g.\ in less than 0.1 seconds.
Other correctness theorems take several minutes.
The slow times seem mainly due to case splits that occur
when dealing with code with several conditionals.
The granularity of the correctness proofs generated by ATC
is at the function and loop level,
i.e.\ there is one theorem per C function or loop,
proved by symbolically executing the function or loop as a whole,
and matching that with the representing C functions.
This process may be slow even for relatively small C functions and loops,
if they have enough conditionals.
An approach to overcome this problem is discussed in \secref{future}.

The hypotheses in the correctness theorems generated by ATC
must be fulfilled by external C code that calls the generated C code,
for the correctness guarantees to hold.
This does not apply to the \code{limit} hypotheses,
which just serve to show that execution terminates,
and which are easily satisfied with sufficiently large values of \code{limit},
which are not part of the data manipulated by the C code
(they are an artifact of the model).
The satisfaction of the hypotheses about the types of the inputs
may be checked by C compilers.
Satisfying the remaining hypotheses
is the responsibility of the calling code;
these are the non-type portions of the guards
(e.g.\ that certain values are in certain ranges),
and the fact that all the arrays are disjoint.
Even though the formal model puts all the arrays in the heap,
the arrays passed to the C functions generated by ATC
could be allocated in the stack (by the callers) as well;
the heap of the formal model more generally represents
externally allocated memory.


\section{Future Work}
\label{future}

An obvious direction of future work is
to add support for increasingly larger subsets of C.
Although the current subset can represent some interesting programs,
many other interesting programs cannot be represented.
Work is underway to add support for structures,
which are commonly used in C.
The main challenge in supporting additional C features
is perhaps the definition of their representations in the shallow embedding.
Given that ACL2 is a very different language from C,
defining such representations may require some thought.
In particular, representing aliasing may involve some complexity,
because of the likely need to explicate some graph structure
that captures aliased data and that is passed through the ACL2 functions.
Handling tree-shaped data without aliasing is comparatively simple,
along the lines of the current treatment of arrays,
which may be generalized to includes structures,
possibly mutually nested with arrays.

Both shallow and deep embedding currently comply to the C18 standard.
However, some C implementations may extend the well-definedness of C constructs
(e.g.\ they may ensure that signed arithmetic is always well-defined,
as two's complement wrap-around as in the x86 processor),
and some applications may rely on this extended well-definedness.
The plan is to parameterize both shallow and deep embedding
over certain implementation-defined characteristics,
captured via ACL2 constrained functions;
different instantiations of these parameters
can be captured via hypotheses over these constrained functions,
and these hypotheses can be used in the applications that require them.
In particular, the behavior of an arithmetic operation
over operands of certain types
whose exact result is not representable in the result type
can be captured via a constrained function.
For C code strictly compliant to the C18 standard,
this constrained function can be hypothesized to return an error indication;
for C code tailored to an implementation based on x86 as above,
this constrained function can be hypothesized to return
the two's complement wrapped-around result.
The same approach is planned for the sizes of integer types,
as alluded to in \secref{shallow}.

The proof (in)efficiency issues discussed in \secref{proofgen}
must be addressed soon.
While there may be ways to mitigate the case splits
by tweaking the symbolic execution,
a more promising approach is to generate finer-grained proofs,
at the level of blocks, or even statements and expressions,
as ATC generates these syntactic entities.
The finer-grained proofs will still use symbolic execution,
but in a more controlled and efficient way, avoiding case splits.
It should be possible to generate finer-grained proofs
that are processed in linear time over the size of the C code.

ATC could be extended with code generation modes
based on direct shallow embedding and direct deep embedding,
analogously to ATJ
\cite{atj-deep,atj-shallow} \citeman{JAVA____ATJ}{java::atj}.
This may require the use of a garbage collector for C.


\section{Related Work}
\label{related}

ATC's shallow embedding of C in ACL2
is similar to the shallow embedding of RAC (Restricted Algorithmic C) in ACL2
\cite[Chapter 15]{russinoff-book}.
Despite their different purposes (code generation vs.\ code verification),
they share the concern of representing C or RAC code in ACL2.

ATJ \cite{atj-deep,atj-shallow} \citeman{JAVA____ATJ}{java::atj}
is a Java code generator for ACL2
primarily based on direct deep embedding and direct shallow embedding.
It also features partial support for inverse shallow embedding,
which was in fact pioneered in ATJ,
and then developed in full in ATC.
While ATJ does not generate proofs yet,
ATC has been generating them from the outset.
While ATC requires the ACL2 code translated to C
to be in a very restricted form,
ATJ translates to Java a much larger subset of ACL2.

Several other theorem provers include code generation facilities
\cite{coq-refman,isa-codegen,pvs-codegen}.
These are based on direct shallow embedding,
which is quite different from ATC's inverse shallow embedding.
Furthermore, the ACL2 language is quite different from those provers' languages,
which are higher-order and strongly typed.
Thus, not many of the ideas from those provers' code generation facilities
may be relevant to ATC.
The C code generator for PVS \cite{pvs-codegen}
may contain the most relevant ideas for ATC,
but likely more for a future code generation mode
based on direct shallow embedding.

Several formalizations of C exist
\cite{c-asm,c-coq,c-hol,c-k,c-papaspyrou}.
These may contain ideas relevant to
extending and improving the formalization of C in ACL2 described in this paper,
which the proofs generated by ATC are based on.


\section*{Acknowledgements}

Thanks to
Ruben Gamboa,
Karthik Nukala, and
Eric Smith
for using the C code generator
and for providing valuable suggestions that led to improvements to the tool.
Thanks to Eric Smith and Karthik Nukala
for developing new APT transformations tailored to the C code generator.
Thanks to David Hardin for
useful discussions on C code generation and related topics.


\bibliographystyle{eptcs}
\bibliography{paper}


\end{document}